\documentclass[twocolumn]{article}

\makeatletter 
\def\@normalsize{\@setsize\normalsize{10pt}\xpt\@xpt
\abovedisplayskip 10pt plus2pt minus5pt\belowdisplayskip \abovedisplayskip
\abovedisplayshortskip \z@ plus3pt\belowdisplayshortskip 
6pt plus3pt minus3pt\let\@listi\@listI}
\def\subsize{\@setsize\subsize{12pt}\xipt\@xipt}
\def\section{\@startsection {section}{1}{\z@}
	{1.8ex plus 1ex minus .2ex} 
	{1.2ex plus .2ex \@afterindentfalse}
	{\large\bf}}
\def\subsection{\@startsection {subsection}{2}{\z@}
	{1.3ex plus 1ex}
	{.8ex plus .2ex \@afterindentfalse}
	{\subsize\bf}}
\def\paragraph{\@startsection {paragraph}{4}{\z@}
	{1.8ex plus .3ex}
	{-1em  \@afterindentfalse}
	{\normalsize\bf}}

\newcommand{\BI}{\begin{itemize}}
\newcommand{\EI}{\end{itemize}}

\usepackage{babel}
\usepackage{url}
\usepackage[hidelinks]{hyperref}
\usepackage{graphicx}

\usepackage{booktabs}    
\usepackage{multirow}    
\usepackage{array}       
\usepackage{caption}     

\makeatletter 
\def\@normalsize{\@setsize\normalsize{10pt}\xpt\@xpt
\abovedisplayskip 10pt plus2pt minus5pt\belowdisplayskip \abovedisplayskip
\abovedisplayshortskip \z@ plus3pt\belowdisplayshortskip 
6pt plus3pt minus3pt\let\@listi\@listI}
\def\subsize{\@setsize\subsize{12pt}\xipt\@xipt}
\def\section{\@startsection
	{section}		
	{1}
	{\z@}
	{2.2ex plus 1ex minus .2ex} 		
	{1.2ex plus .2ex \@afterindentfalse}	
	{\large\bf}}				
\def\subsection{\@startsection
	{subsection}
	{2}
	{\z@}
	{2.0ex plus 1ex}			
	{.8ex plus .2ex \@afterindentfalse}	
	{\subsize\bf}}				
\def\subsubsection{\@startsection
	{subsection}
	{3}
	{\z@}
	{1.8ex plus 1ex}			
	{.8ex plus .2ex \@afterindentfalse}	
	{\normalsize\bf}}			
\def\paragraph{\@startsection 
	{paragraph}
	{4}
	{\z@}
	{1.8ex plus .3ex}			
	{-1em \@afterindentfalse}		
	{\normalsize\bf}}			

\usepackage[a4paper,
    left=2cm,
    right=2cm,
    top=2cm,
    bottom=3.1cm
]{geometry}

\begin{document}

\twocolumn[{
\centering
\Large\bfseries
Challenges in Developing Secure Software - Results of an Interview Study in the German Software Industry\par
\bigskip

\normalsize
\begin{minipage}{0.92\textwidth}\centering
Alex R. Mattukat, Timo Langstrof, Horst Lichter\par
\end{minipage}

\vspace{0.75em}

\small
\begin{minipage}{0.95\textwidth}\centering
RWTH Aachen University, Research Group Software Construction

\end{minipage}
  \small \texttt{\{mattukat, lichter\}@swc.rwth-aachen.de},
  \small \texttt{timo.langstrof@rwth-aachen.de}

\bigskip
}]


\begin{abstract}
\footnote{This paper is an English translation of our paper published in the German ``Softwaretechnik Trends'' journal \cite{Mattukat.2025}. We used AI assistants for the translation. In the first step, we used DeepL to translate the German text into English paragraph-wise; that is, each paragraph was first translated with DeepL, then manually reviewed by the first author, and improved by him if necessary. After the translation of the entire paper was completed, the third author reviewed the complete paper again and made necessary improvements to the translation. Finally, Grammarly was used to make grammatical and stylistic improvements.}
The damage caused by cybercrime makes the development of secure software inevitable. Although many tools and frameworks exist to support the development of secure software, statistics on cybercrime show no improvement in recent years. To understand the challenges software companies face in developing secure software, we conducted an interview study with 19 industry experts from 12 cross-industry companies. The results of our study show that the challenges are mainly due to high complexity, a lack of security awareness, and unsuitable processes, which are further exacerbated by an immediate lack of skilled personnel. This article presents our study and the challenges we identified, and derives potential research directions from them.
\end{abstract}

\section{Introduction}

Software security (hereinafter referred to as “security”) and cybercrime have become central topics in the software industry. Microsoft's latest Digital Defense Report ranks Germany fourth among the countries most frequently targeted by cyberattacks \cite{Microsoft.2025}; in 2024, over 130,000 cases of cybercrime were reported \cite{BKA2025CybercrimeStatista}. The annual costs incurred by cybercrime are also skyrocketing. While the costs in Germany amounted to €179 billion in 2024 \cite{SoSafe2025CybercrimeTrends}, they are estimated to reach almost €900 billion in 2030, and as much as €17 trillion (!) worldwide \cite{Statista.2025.Costs.Ger, Statista.2025.Costs.World}. An analogy is needed to make these sums somewhat tangible. If cybercrime were an independent nation, it would be the third-largest economy in the world after the US and China \cite{Martins.2025}.

To improve this situation, there are many guidelines such as the EU-wide NIS2 Directive \cite{NIS2}, standards and certifications such as ISO/IEC 27001 \cite{ISO.27001}, as well as general approaches such as Security by Design \cite{Waidner.2013}, the Microsoft Secure Development Lifecycle \cite{Microsoft.SDL}, the NIST Cybersecurity Framework \cite{NIST.CSF}, and the OWASP Developer Guide \cite{OWASP.DevGuide}. Nevertheless, the figures and trends relating to cybercrime do not indicate any improvement in the situation. It is therefore worthwhile to understand the challenges that need to be overcome from the perspective of the software industry. We investigated these in an expert interview study.

\section{Methodology of the Interview Study}
The interview guide was designed in accordance with the guidelines by Robson and McCarten \cite{Robson.2016} and Charmaz \cite{Charmaz.2014}. The open nature of the semi-structured interview guide takes into account the diversity of terminology used in the field of security. The guide consists of five thematic blocks, each containing a list of open-ended questions and a block of socio-demographic questions. In this article, we will focus on the thematic block on current and future challenges in the field of security. The guide can be viewed via our GitLab repository\footnote{\url{https://git.rwth-aachen.de/swc-public/public-research-documents/security-interview-study}}.

The target groups for this study were developers, architects, and project managers who deal with security-related issues. In addition, they had to work for a software development company. The target group was deliberately kept very broad, as our study was intended to cover security broadly and enable trend analysis. Participants were selected using a mix of purposive sampling and snowball sampling.

For purposive sampling, we contacted individuals from our professional networks via LinkedIn or email. The request contained background information on the study, an invitation to participate, and, for the purpose of snowball sampling, a request to forward it to other potentially interested parties. This enabled us to reach 19 participants from 12 companies. Eighteen interviews were conducted individually via online meeting tools, and one was conducted as an on-site double interview at the company. Except for one individual interview in English, all interviews were conducted in German. The companies included small and medium-sized enterprises (SME) ($<1,000$), large enterprises ($>1,000$, $< 10,000$), and corporations ($>10,000$) from the following sectors: financial services, automotive, supply chain management, public transport, energy and environment, and software services. Accordingly, specialists from both highly regulated and little to non-regulated industries were interviewed. Table \ref{table} provides an overview of the interviews. The table indicates the interviewee's role (manager, architect, developer), the company's industry and size; the interviews are grouped by company affiliation.

The transcripts were coded using an iterative, hybrid inductive-deductive coding approach, following Mayring \cite{Mayring2014QualitativeContentAnalysis}. Two authors performed the coding independently. In the first iteration (January 6 to May 12, 2025), an initial code book (inductive) was created based on a random 25\% sample, the results obtained were compared, deviations were discussed, and finally a consolidated code book was created. Subsequently, all transcripts were coded deductively. Missing or inappropriate codes were supplemented or corrected jointly. In the second iteration (June 1 to September 15, 2025), both authors reviewed the refined codebook again using another random 15\% sample and adjusted it were necessary. Then, both authors recoded all transcripts again deductively using the final codebook. In a last step, “focused coding” was performed according to Saldaña \cite{Saldana.2016}. To this end, both authors jointly analyzed the codes and transcripts to identify differences, similarities, and inconsistencies. The challenges identified from the codes were also extracted, and the frequency of occurrence was determined.

\begin{table}[tbh] 
\centering
\caption{Overview of interview participants grouped by company affiliation}
\begin{tabular}{@{}llll@{}} 
\toprule 
\textbf{ID} & \textbf{Role} & \textbf{Sector} & \textbf{Size} \\
\midrule 
\specialrule{0.1pt}{0.1pt}{0.1pt}

$I_{1}$  & Manag.    & \multirow{3}{*}{\begin{tabular}[c]{@{}l@{}}Energy and\\ Environment\end{tabular}}  & \multirow{3}{*}{SME} \\
$I_{2}$\textsuperscript{a}  & Manag.    &                                      &                           \\
$I_{3}$\textsuperscript{a}  & Arch.  &                                      &                           \\ 
\specialrule{0.1pt}{0.1pt}{0.1pt}

$I_{4}$  & Dev. & \multirow{3}{*}{\begin{tabular}[c]{@{}l@{}}Supply Chain\\ Management\end{tabular}}        & \multirow{3}{*}{Corporation} \\
$I_{5}$  & Manag.    &                                      &                              \\
$I_{6}$  & Manag.    &                                      &                              \\
\specialrule{0.1pt}{0.1pt}{0.1pt}

$I_{7}$  & Arch.  & \multirow{2}{*}{\begin{tabular}[c]{@{}l@{}}Financial-\\ Services\end{tabular}}            & \multirow{2}{*}{Large Enterp. }   \\
$I_{8}$  & Dev. &                                      &                              \\
\specialrule{0.1pt}{0.1pt}{0.1pt}

$I_{9}$  & Manag.    & \multirow{2}{*}{Software Services} & \multirow{2}{*}{SME}         \\
$I_{10}$ & Manag.    &                                      &                              \\
\specialrule{0.1pt}{0.1pt}{0.1pt}

$I_{11}$ & Dev. & \multirow{2}{*}{Public Transport}   & \multirow{2}{*}{SME}     \\

$I_{12}$ & Arch.  &                                      &                              \\
\specialrule{0.1pt}{0.1pt}{0.1pt}

$I_{13}$ & Arch.  & Automotive                            & Corporation                      \\
\specialrule{0.1pt}{0.1pt}{0.1pt}

$I_{14}$\textsuperscript{b} & Arch.  & Software Services                  & Corporation                \\
\specialrule{0.1pt}{0.1pt}{0.1pt}

$I_{15}$ & Manag.    & IT Infrastructure                        & SME                        \\
\specialrule{0.1pt}{0.1pt}{0.1pt}

$I_{16}$ & Manag.    & Automotive                            & Corporation                      \\
\specialrule{0.1pt}{0.1pt}{0.1pt}

$I_{17}$ & Manag.    & Software Services                  & SME                         \\
\specialrule{0.1pt}{0.1pt}{0.1pt}

$I_{18}$ & Manag.    & Software Services                  & Large Enterp.                 \\
\specialrule{0.1pt}{0.1pt}{0.1pt}

$I_{19}$ & Arch.  & Software Services                             & Corporation                  \\
\bottomrule 
\end{tabular}

\vspace{1ex}
\raggedright
\footnotesize
\textsuperscript{a}on-site- double interview ; 
\textsuperscript{b}interview in English
\label{table}
\end{table}

\section{Results}
\begin{figure*}[t]
    \centering
    \includegraphics[width=\textwidth]{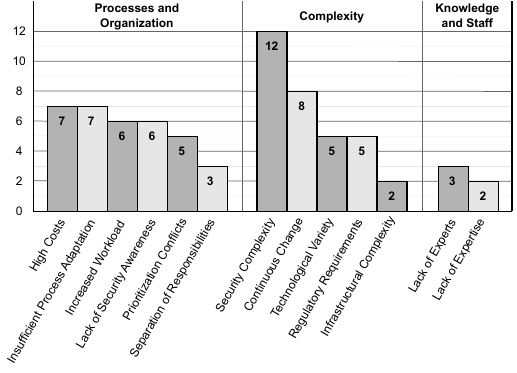}
    \caption{Identified challenges, sorted by their category}
    \label{fig:quantitativ}
\end{figure*}

Based on our study, we were able to identify 13 challenges, which we divided into three categories: challenges relating to project- or company-wide \textbf{processes and organization}, those arising from the high level of \textbf{complexity} associated with security, and those caused by a lack of \textbf{knowledge and personnel}. Challenges in the “processes and organization” category were identified most frequently (34 mentions). Challenges related to complexity were mentioned with similar frequency (32 mentions). Challenges due to a lack of knowledge and personnel were mentioned least frequently (5 mentions). Figure \ref{fig:quantitativ} shows these categories and the frequency with which the challenges were mentioned. In the following, we describe these challenges (the number of mentions is given in brackets) and provide a representative excerpt from a transcript. The transcripts were translated from German using DeepL version 25.10.3, including manual review and, where necessary, corrections by an author. Since participants expressly asked us not to publish the transcripts despite sufficient anonymization, we cannot make them available in full.

\subsection{Processes and Organization}
This category includes procedural and organizational challenges that make it difficult to develop secure software.

\paragraph{High Costs (7):}
The increased costs in the development, testing, and operation of software caused by the implementation of security measures were most frequently cited as a challenge. These costs can lead to prioritization conflicts, potentially resulting in reduced or postponed security-related activities.

\noindent \textit{$I_{5}$: ``The most difficult thing, in my opinion, is that—if you take it really seriously—it becomes quite expensive, because it can be quite complex. And that you really have to find the right balance between costs and truly necessary security.''}

\paragraph{Insufficient Process Adaptation (6):}
Missing or inadequate processes for the systematic implementation and continuous improvement of security were similarly often mentioned. In many cases, there are only a few processes in place to firmly integrate security aspects into software development.

\noindent \textit{$I_{16}$: ``Everything that was done on-premises is attempted to be done in the cloud as well, and sometimes it is extremely challenging, if not completely impossible, to adhere to.''}

\paragraph{Increased Workload (6):}
The integration of security into development and operational processes adds a considerable workload, as it requires additional coordination, documentation, testing procedures, and validation steps, among other things. This challenge is closely related to the challenge of the lack of experts, which will be discussed later.

\noindent \textit{$I_{9}$: ``People already have so much to do. They have to help define requirements, understand the technical aspects, implement the functionality, document it, write test cases for it — unit tests, integration tests, end-to-end tests — and then there's security too. That's a lot to take on.''}

\paragraph{Lack of Security Awareness (6):}
Both within development departments and among customers, there is often a lack of understanding of the relevance and necessity of considering security as a key issue. Among other things, this leads to conflicts of priority between security and functionality, a ``silo mentality'', and financing problems in projects. In addition, the human factor was identified as a weak point: misconduct, inadequate training, and insufficient awareness increase the risk of security incidents.

\noindent \textit{$I_{9}$: ``Awareness must be created. This is something that needs to be actively addressed. And this is actually the first and biggest hurdle.''}

\paragraph{Prioritization Conflicts (5):}
Prioritization conflicts between implementing security requirements and other requirements, particularly functional requirements, were also mentioned several times. Security measures are sometimes perceived as a limitation to functionality or other quality attributes and must therefore be reconciled with other development goals.

\noindent \textit{$I_{5}$: ``The customer never says, ``I do not need that''. They are not that brave. But when it comes to prioritization, it's unfortunately often the case that features are given higher priority.''}

\paragraph{Separation of Responsibilities (3):}
The organizational separation of responsibilities into separate departments leads to inefficient communication of knowledge and experience regarding security solutions. This lack of communication hampers the holistic and sustainable implementation of security in companies.

\noindent \textit{$I_{19}$: ``The customer's policy is that anything that does not need to be known by others is not disclosed. Unfortunately, this means that many of these best practices are not properly shared by the teams implementing them.''}

\subsection{Complexity}
The second category includes challenges caused by high technical or structural complexity and that are related to security, which make it more difficult to develop secure software.

\paragraph{Security Complexity (12):}
The complexity of security as a whole was mentioned most frequently. Security is perceived as a complex and multifaceted issue. It encompasses numerous aspects, technologies, and responsibilities whose interactions are difficult to oversee and control.

\noindent \textit{$I_{5}$: ``The big challenge is that it is so complex and has so many aspects. As a developer, you cannot do everything on the side as a secondary activity.''}

\paragraph{Continuous Change (8):}
The rapid pace of innovation, new technologies, and constantly changing threats and regulatory requirements pose a further challenge. They complicate efforts to keep pace with the development of appropriately secure software. Security cannot be established as a one-time measure; it requires continuous adaptation and maintenance at both the technical and organizational levels. This constant change increases complexity and challenges companies to regularly update processes, tools, and responsibilities to ensure an adequate level of security.

\noindent \textit{$I_{9}$: ``At the same time, I see this as an ongoing challenge. It is not a task that you tackle once with a huge effort and then put aside and say, “Okay, now it's working.” Rather, it is something that needs to be continuously developed.''}

\paragraph{Technological Variety (5):}
The variety of technical solutions, frameworks, and tools in the field of security sometimes causes difficulties for companies to apply them in a targeted and consistent manner. This applies particularly to selecting suitable technologies and integrating them into existing architectures and infrastructures.

\noindent \textit{$I_{8}$: ``It starts at the bottom with Kubernetes and goes all the way to the front end, where there is such a wide spectrum. I think the challenge lies in maintaining an overview and finding tools to help.''}

\paragraph{Regulatory Requirements (5):}
In many cases, regulatory or legal requirements must be considered in software development. On the one hand, the multitude, inconsistency, and sometimes unclear wording of regulatory requirements make it difficult to fully implement security requirements consistently. On the other hand, regulatory requirements complicate the software development process, thereby affecting the implementation of other, non-security-related requirements.

\noindent \textit{$I_{7}$: ``The challenge is that we want to offer customers more. But due to these conditions, due to our 2,000 pages of security guidelines, you cannot be that creative. You have to pay attention to everything!''}

\paragraph{Infrastructural Complexity (2):}
Another challenge lies in the complexity of the infrastructure. Different deployment environments and service models, such as hybrid cloud environments, require tailored security measures. This increases and complicates the administrative effort.

\noindent \textit{$I_{14}$: ``Depending on your infrastructure landscape, there are best practices how to organize security boundaries among your infrastructure. After that, it's important to have sufficient knowledge in understanding different vectors of attacks, what are nowadays the most popular security vulnerabilities, and you need to constantly monitor this information and react quickly.''}

\subsection{Knowledge and Staff}
The third category includes challenges arising from a lack of experts and expertise.

\paragraph{Lack of Experts (3):}
The lack of experts was explicitly mentioned in three interviews and implicitly addressed in others, but this is not reflected in the explicit number of mentions (3). This lack not only affects the development of secure software but also profoundly impacts project organization and software development in general. It often results in employees without the necessary knowledge being responsible for security-related activities. This usually results in these employees being severely overworked and thus no longer available to the extent required for their original areas of responsibility.

\noindent \textit{$I_{18}$: ``In the end, we do not have a single person in our company who is an expert in this field at the moment. These are all things we have learned on the side. And it would probably be a good idea to have someone who deals with this full-time, who is part of the team and who, for example, makes sure that we take these things into account during development.''}

\paragraph{Lack of Expertise (2):}
As already mentioned, due to a lack of experts in the security sector, employees without the relevant qualifications or training are often assigned to security-related tasks. A lack of specialist expertise is therefore perceived as a key challenge, as it impairs both the efficiency of development and the quality of the software developed.

\noindent \textit{$I_{5}$: ``We first need to build up experience and will probably also need to purchase it externally.''}

\section{Discussion}
Our findings show that the challenges of developing secure software are both technical and organizational.

Projects and companies often lack the appropriate processes and security awareness to integrate security across all aspects of the software development process. As a result, the workload associated with developing secure software continues to increase. The lack of experts further exacerbates the situation, as employees without the relevant security expertise are assigned to these tasks.

In addition to the increased workload, financing the projects poses another challenge. The low level of security awareness, especially among management and customers, makes financing difficult. Security is often perceived as an additional, cost-driving task rather than an integral and necessary part of software and software development. This, in turn, leads to prioritization conflicts between the implementation of functional and security requirements.

In addition to organizational challenges, implementing security measures increases complexity. It became clear that respondents perceive dealing with security in general and developing secure software as difficult and complex. The reasons for this complexity are multifaceted. In some cases, they can be attributed to technical aspects such as the multitude of security threats and the constant evolution of technologies. However, the reasons for the increased complexity are often difficult to identify. The increased complexity arises from the interplay of technical challenges, such as integrating appropriate security mechanisms and responding to evolving threats, and organizational factors, such as inadequate processes, a shortage of experts, and poor communication among involved roles and departments.

The complexity involved in developing and operating secure software poses major challenges for companies. These challenges cannot be overcome by technical measures alone, but also require changes to organizational structures, clear responsibilities, and an appropriate security culture. Suitable technologies and best practices are often available, but are not always effectively integrated into the development process.

Implementing the security-by-design approach promises to improve this situation by integrating security early and systematically into the development process \cite{Waidner.2013}. Frameworks for systematic implementation already exist \cite{Microsoft.SDL, NIST.CSF, OWASP.DevGuide}, but a lack of security awareness often makes this difficult. Without this security awareness, it is difficult—if not impossible—to develop and promote the security culture necessary for software development in projects and companies. For companies to develop secure software efficiently and effectively, they need an initial impetus to establish security awareness and accountability. Of course, this costs time and money, as any change does. However, the effort required to establish a security culture and develop customized processes is essential due to the rise in cybercrime and pays off in the long run. The following transcript excerpt succinctly summarizes this discussion:
 
\noindent \textit{$I_{19}$: ``Often enough, it is simply awareness, and above all awareness from the very beginning. Adding security later is something that simply does not work. It has to be there from the start.''}

\subsection{Threats to Validity}
This study is based on 19 interviews with specialists from twelve German software companies. Although the sample is intended to ensure thematic breadth, its size does not allow for statistical generalization of the results or for proof of causal relationships. Rather, the aim of the study was to identify the challenges perceived within the companies. Another source of bias lies in participants' subjective perceptions. The experiences described are based on individual perspectives and specific organizational contexts and may not be transferable to other companies. The interpretation of the data is also subject to certain limitations. Despite the hybrid coding process by two authors and the joint development of a consolidated codebook, a certain degree of subjectivity in the categorization of challenges remains unavoidable. Finally, it should be noted that the regulatory and technological framework is constantly changing, so the results of this study represent a snapshot of the years 2023 to 2025.

\section{Outlook}
Our study results offer some starting points for further research. For example, a study could examine the relationships between organizational maturity, security awareness, and technical implementation. This could provide insights for deriving targeted measures to improve security practices in software development projects. A better understanding of these relationships could also help develop effective strategies to promote a company-wide security culture and facilitate the integration of security measures into existing development processes. A comparative study of different company sizes and industries could also be valuable in identifying industry-specific patterns and success factors.

In addition, the challenges arising from increased complexity provide a basis for future work to better understand complexity in the context of security and make it more manageable. This could lead to the development of approaches that comprehensively consider the technical and organizational aspects of security and systematically analyze their interactions. Future research should also aim to develop processes, methods, and tools that enable developers and architects to incorporate security aspects into development processes more efficiently and consistently.

Further insights could be provided by empirical studies on the effectiveness of specific security-enhancing measures, such as awareness programs, process adjustments, or the introduction of security-by-design approaches. Based on such studies, practical recommendations could be developed to address the challenges identified in this article.

\section*{Acknowledgements}
We would like to express our sincere thanks to all participants in this study for their cooperation, openness, and time invested.

\bibliographystyle{IEEEtran}
\bibliography{Bibliography}

@misc{BKA2025CybercrimeStatista,
  author       = {{Bundeskriminalamt}},
  title        = {{Number of cases of cybercrime recorded by police in Germany from 2007 to 2024 [Graph]}},
  year         = {2025},
  month        = apr,
  howpublished = {In Statista},
  note         = {Retrieved October 9, 2025, from \url{https://www.statista.com/statistics/1360141/cyber-crime-cases-recorded-police-germany/}},
}

@electronic{Statista.2025.Costs.Ger,
 author = {{Statista}},
 organization = {{Statista}},
 year = {2025},
 title = {{Cybersecurity - Germany}},
 url = {https://www.statista.com/outlook/tmo/cybersecurity/germany?currency=EUR},
 note = {Last accessed: 13.10.2025}
}

@electronic{Statista.2025.Costs.World,
 author = {{Statista}},
 organization = {{Statista}},
 year = {2025},
 title = {{Cybersecurity - Worldwide}},
 url = {https://www.statista.com/outlook/tmo/cybersecurity/worldwide?currency=EUR},
 note = {Last accessed: 13.10.2025}
}

@electronic{Microsoft.SDL,
 author = {{Microsoft}},
 organization = {{Microsoft}},
 year = {2025},
 title = {{Security Development Lifecycle (SDL)}},
 url = {https://www.microsoft.com/en-us/securityengineering/sdl},
 note = {Last accessed: 14.10.2025}
}

@electronic{NIST.CSF,
 author = {{NIST}},
 organization = {{National Institute of Standards and Technology (NIST)}},
 year = {2025},
 title = {{Cybersecurity Framework}},
 url = {https://www.nist.gov/cyberframework},
 note = {Last accessed: 14.10.2025}
}

@electronic{OWASP.DevGuide,
 author = {{OWASP}},
 organization = {{Open Worldwide Application Security Project (OWASP)}},
 year = {2025},
 title = {{OWASP Developer Guide}},
 url = {https://owasp.org/www-project-developer-guide/},
 note = {Last accessed: 14.10.2025}
}

@techreport{Waidner.2013,
  title         ={{Entwicklung sicherer Software durch Security by Design}},
  author        ={Waidner, Michael and Backes, Michael and M{\"u}ller-Quade, J{\"o}rn and others},
  institution  = {{Fraunhofer-Institut für sichere Informationstechnologie}},
  year          = {2013},
  publisher     = {Fraunhofer Verlage},
  issn          = {2192-8169},
  isbn          = {978-3-8396-0567-7}
}

@article{Martins.2025,
 author = {Martins, Ant{\'o}nio Miguel and Moutinho, Nuno and Cr{\'o}, Susana},
 year = {2025},
 title = {Stock market effects of major cyber-attacks: evidence for breached and cybersecurity listed firms},
 issn = {1750-2071},
 journal = {Journal of Banking Regulation},
 doi = {10.1057/s41261-025-00293-y}
}

@book{Robson.2016,
 author = {Robson, Colin and McCartan, Kieran},
 year = {2016},
 title = {Real World Research: A Resource for Users of Social Research Methods in Applied Settings},
 edition = {4th},
 publisher = {{John Wiley {\&} Sons}},
 isbn = {9781119144854}
}

@book{Charmaz.2014,
 author = {Charmaz, Kathy},
 year = {2014},
 title = {Constructing Grounded Theory},
 edition = {2},
 publisher = {{SAGE Publications}},
 isbn = {978-0-85702-9133}
}

@misc{NIS2,
  title        = {{Richtlinie (EU) 2022/2555 des Europäischen Parlaments und des Rates vom 14. Dezember 2022 über Maßnahmen für ein hohes gemeinsames Niveau der Cybersicherheit in der Union (NIS 2-Richtlinie)}},
  howpublished = {{Amtliche Reihe des Amtsblatts der Europäischen Union, L 333, 27. Dezember 2022, S. 80–152, Deutsch}},
  year         = {2022}
}

@misc{ISO.27001,
 year = {2023},
 title = {{EN ISO/IEC 27001:2023 - Information security, cybersecurity and privacy protection - Information security management systems - Requirements (ISO/IEC 27001:2022)}},
 number = {{EN ISO/IEC 27001:2023}},
 author = {{International Organization for Standardization and International Electrotechnical Commission}}
}

@book{Saldana.2016,
 author = {Salda{\~n}a, Johnny},
 year = {2016},
 title = {The Coding Manual for Qualitative Researchers},
 publisher = {{SAGE Publications}},
 isbn = {978-1-4739-0248-0}
}

@book{Mayring2014QualitativeContentAnalysis,
  author    = {Philipp Mayring},
  title     = {Qualitative content analysis: theoretical foundation, basic procedures and software solution},
  year      = {2014},
  address   = {Klagenfurt},
  url       = {https://nbn-resolving.org/urn:nbn:de:0168-ssoar-395173}
}

@techreport{Microsoft.2025,
  author       = {{Microsoft}},
  title        = {Microsoft Digital Defense Report 2025},
  institution  = {Microsoft Corporation},
  year         = {2025},
  url          = {https://cdn-dynmedia-1.microsoft.com/is/content/microsoftcorp/microsoft/msc/documents/presentations/CSR/Microsoft-Digital-Defense-Report-2025.pdf},
  note         = {Accessed: 14 October 2025}
}

@techreport{SoSafe2025CybercrimeTrends,
  title        = {{Cybercrime-Trends 2025}},
  institution  = {{SoSafe SE}},
  address      = {{Köln, Deutschland}},
  year         = {2025},
  type         = {Whitepaper},
  url          = {https://sosafe-awareness.com/de/ressourcen/reports/cybercrime-trends/},
  urldate      = {2025-11-05},
  language     = {German}
}

@article{Mattukat.2025,
  author    = {Alex R. Mattukat and Timo Langstrof and Horst Lichter},
  title     = {{Herausforderungen bei der Entwicklung sicherer Software: Ergebnisse einer Interviewstudie in der deutschen Softwareindustrie}},
  journal   = {Softwaretechnik–Trends},
  volume    = {45},
  number    = {4},
  year      = {2025},
  pages     = {2--7},
  issn      = {0720-8928},
}

\end{document}